\documentclass[letterpaper,12pt]{article}
\usepackage{graphicx}
\usepackage{epstopdf}
\usepackage{amsmath}
\usepackage{amssymb}
\usepackage{amsthm}
\usepackage{ccaption}
\usepackage[usenames]{color}
\usepackage{cite}
\usepackage{epsf}
\def\TL{\hfil$\displaystyle{##}$}
\def\TR{$\displaystyle{{}##}$\hfil}

\def\comment#1{}
\def\fixit#1{}

\def\overleftrightarrow#1{\vbox{\ialign{##\crcr
     $\leftrightarrow$\crcr\noalign{\kern-0pt\nointerlineskip}
     $\hfil\displaystyle{#1}\hfil$\crcr}}}

\def\lsim{\mathrel{\mathstrut\smash{\ooalign{\raise2.5pt\hbox{$<$}\cr\lower2.5pt\hbox{$\sim$}}}}}
\def\gsim{\mathrel{\mathstrut\smash{\ooalign{\raise2.5pt\hbox{$>$}\cr\lower2.5pt\hbox{$\sim$}}}}}
\def\sqr#1#2{{\vcenter{\vbox{\hrule height.#2pt
         \hbox{\vrule width.#2pt height#1pt \kern#1pt
            \vrule width.#2pt}
         \hrule height.#2pt}}}}

\def\href#1#2{#2}  

\def\lbldef#1#2{\expandafter\gdef\csname #1\endcsname {#2}}
\def\eqn#1#2{\lbldef{#1}{(\ref{#1})}%
\begin{equation} #2 \label{#1} \end{equation}}
\def\eqalign#1{\vcenter{\openup1\jot
    \halign{\strut\span\TL & \span\TR\cr #1 \cr
   }}}

\makeatletter
\@addtoreset{equation}{section}

\makeatletter
\renewcommand\section{\@startsection {section}{1}{\z@}%
                                   {-3.5ex \@plus -1ex \@minus -.2ex}%nn
                                   {2.3ex \@plus.2ex}%
                                   {\normalfont\large\bfseries}}
\renewcommand\subsection{\@startsection{subsection}{2}{\z@}%
                                     {-3.25ex\@plus -1ex \@minus -.2ex}%
                                     {1.5ex \@plus .2ex}%
                                     {\normalfont\bfseries}}

\def\baselinestretch{1.2}
\def\[{\left [}
\def\]{\right ]}
\def\({\left (}
\def\){\right )}

\def\b{\beta}
\def\om{\omega}

\def\vp{\varphi}
\def\d{\delta}

\def\CM{{\cal M}}
\def\CO{{\cal O}}

\def\CL{{\cal L}}

\def\S{{\bf S}}

\def\p{\partial}

\def\ket#1{\mid \! #1 \, \rangle}
\def\state#1{\mid \! #1 \, \rangle}

\def\expval#1{{\langle \, #1  \, \rangle}}

\newcommand{\be}{\begin{equation}}
\newcommand{\ee}{\end{equation}}
\newcommand{\bea}{\begin{eqnarray}}
\newcommand{\eea}{\end{eqnarray}}

\def\A5S5{AdS$_{5}$ $\times$ $\S^5$}

\def\Cther{\mathcal{O}_{thermal}}

\def\mdiscr{\mathcal{M}_{bas}}
\def\mcont{\mathcal{M}_{sup}}

\newcommand{\ads}[1]{{\rm AdS}_{#1}}

\newcommand{\bbibitem}[1]{\bibitem{#1}\marginpar{#1}}

\def\Label#1{\label{#1}%
  \smash{\hbox to0pt{\raise1ex\hbox{\tiny[#1]}\hss}}}
\def\noLabels{\let\Label=\label}
\def\nobbibitem{\let\bbibitem=\bibitem}

\title{{\bf  Typicality versus thermality: An analytic distinction}}

\author{Vijay Balasubramanian$^\sharp$, Bart{\l}omiej Czech$^\sharp$, Veronika E Hubeny$^\dagger$,\\
 Klaus Larjo$^\sharp$, Mukund Rangamani$^\dagger$ and Joan Sim{\'o}n$^{\sharp,\flat}$ 
 \footnote{vijay@physics.upenn.edu, czech@sas.upenn.edu, veronika.hubeny@durham.ac.uk, \newline  klarjo@physics.upenn.edu, mukund.rangamani@durham.ac.uk, JSimonSoler@lbl.gov}
 \\ [1mm]
\small \sl $^\sharp$David Rittenhouse Laboratories, University of Pennsylvania,
\\[-1.5mm]
\small \sl Philadelphia, PA,  19104, USA\\
\small \sl $^\dagger$ \; Centre for Particle Theory \& Department of
Mathematical Sciences,
\\[-1.5mm]
\small \sl Science Laboratories, South Road, Durham DH1 3LE, United Kingdom \\
\small \sl $^\flat$ \; Dept of Physics \& Theoretical Physics Group, LBNL
\\[-1.5mm]
\small \sl University of California, Berkeley,  94720, USA \\
}

\begin{document}

\noLabels % uncomment for final production
\nobbibitem % uncomment for final production

\setlength{\baselineskip}{16pt}
\begin{titlepage}
\maketitle
\begin{picture}(0,0)(0,0)
\put(350, 420){UPR-T-1170, DCPT-07/01}
\put(350,403){UCB-PTH-07/01, LBNL-62258}
\put(350,386){NSF-KITP-07-02}
\put(350, 370){hep-th/0701122}
\end{picture}
\vspace{-36pt}

%Abstract
\begin{abstract}
In systems with a large degeneracy of states such as black holes, one expects that the average value of probe correlation functions will be well approximated by the thermal ensemble. To understand how correlation functions in individual microstates differ from the canonical ensemble average and from each other, we study the variances in correlators.  Using general statistical considerations, we show that the variance between microstates will be exponentially suppressed in the entropy.   However, by exploiting the analytic properties of correlation functions we argue that these variances are amplified in imaginary time, thereby distinguishing pure states from the thermal density matrix.  We demonstrate our general results in specific examples and argue that our results apply to the microstates of black holes.

\end{abstract}
\thispagestyle{empty}
\setcounter{page}{0}
\end{titlepage}

\renewcommand{\baselinestretch}{1.2}  %looks better
%_________________________________________

%\tableofcontents

%~~~~~~~~~~~~~~~~~~~~~~~~~~~~~~~~~~~~~~~~~~~~~~~
\section{Introduction}
\label{intro}
%~~~~~~~~~~~~~~~~~~~~~~~~~~~~~~~~~~~~~~~~~~~~~~~

Our usual understanding of effective field theory suggests that
the well-known classical geometry of a  black hole spacetime is
valid past the horizon  through to a Planck distance away from the
singularity.   Recently, however, it has been proposed that the
microstates of black holes form a sort of  spacetime foam that
extends throughout the interior of the horizon \cite{fuzzball}.   Such a
magnification of the characteristic length scale of quantum
gravity recalls macroscopic manifestations of quantum mechanics
such as Bose condensation that appear due to the statistical
interactions of many microscopic constituents.

The proposed non-singular, horizon-free, spacetime foam
microstates have only been constructed for certain extremal black
holes in string theory.\footnote{The conjectures of \cite{fuzzball} and results cited there mostly concern extremal black holes of vanishing area.   The half-BPS extremal black hole of $\ads{5}$ has been similarly analyzed in \cite{llm,babel}.    A large set of candidate microstates for asymptotically flat finite area black holes in five and four dimensions was constructed in \cite{finitearea}.}    In the classical limit, they are
characterized by a large topologically complex region within which the
characteristic scale is microscopic.      However,
the foam itself extends out to macroscopic distances, behaving in
some cases like an incompressible fluid.  These solutions should
be understood as a classical moduli space of candidate microstates
which must be quantized.   A key question is whether a {\it
semiclassical} observer probing such quantized microstates makes
measurements that are different from those expected from the usual
black hole spacetimes.    If they are not, the usual classical
geometries are the correct effective description of a black hole
for a semiclassical observer, even though observers with more
precise tools or greater patience might observe a horizon- and
singularity-free universe \cite{vijaydoninteg,coarsegrain1,coarsegrain2}.

These questions are most efficiently studied when the black holes
and their microstates are asymptotic to AdS spacetime.    In this
case, the dual field theory can be used to enumerate the
microstates and quantize them.   We can then ask whether the
correlation functions of probe operators can distinguish the
microstates from each other or  tell them apart from a mixed
state.      This  involves computing the variance of 
correlation functions over a suitable ensemble of pure microstates
with given macroscopic quantum numbers.    In Sec.~2 we discuss
different natural notions of variance and  show that, if the
entropy associated to the ensemble of microstates is $S$, the
variance over these microstates of any local correlation function
will be suppressed by a factor of $e^{-S}$ (relative to a natural ensemble of the basis microstates).   This result, which
applies equally to free and interacting theories,\footnote{Free
theories typically have a highly degenerate spectrum with large
gaps, unlike the interacting theories describing black holes, which
typically have non-degenerate spectra with gaps that scale with
$e^{-S}$.  In both cases we imagine some energy resolution $\Delta
E$ of the macroscopic measuring device and states falling within
this resolution contribute to the entropy.} occurs for statistical
reasons -- almost all microstates are statistically random quantum
superpositions of a basis of states and thus lie very close to a
certain {\it typical} state.    This drives the universality of local
correlators.

To  detect the differences between microstates, an observable must
defeat the $e^{-S}$ suppression of variance.  Since correlation
functions that oscillate in real time grow exponentially in
imaginary time, it is possible that the analytic structure of
correlators could help with separating microstates.    Indeed,
several authors \cite{kos,fhks,leviross,mosheetc, festucciaL,inflateads,maeda,kabat} have suggested that in the AdS/CFT correspondence the structure of correlators in the complex time plane can probe the region behind a horizon and in the
vicinity of a black hole singularity.    Many of these articles  study the physics of an eternal black hole in AdS spacetime.   Such spacetimes have two asymptotic regions, and in the dual description, the vacuum configuration is a pure entangled state in a product of two CFTs, one associated to each asymptotic region \cite{juaneternal}. Tracing over one CFT produces a thermal density matrix. Here we are
examining how the pure underlying microstates, possibly described
by a horizon-free, singularity-free, and
second-asymptotic-region-free spacetime, are described in terms of
a {\it single} CFT. In this context,   we examine how local
correlation functions computed in the underlying pure states
behave in the complex time plane, and extract the timescales at
which they differ significantly from the thermal averages computed from the eternal black hole.

Even within a single CFT we can construct  a thermal density matrix.   What is the dual description of this ensemble in the AdS/CFT correspondence?   Sometimes the dual is taken to be a single geometry, i.e., an eternal black hole.   Our results suggest that the dual is not described in this way by a single geometry, but rather by an ensemble of universes (in a quantum cosmological sense) corresponding to each microstate.   This is because, in both CFT and spacetime, each time a measurement is performed, one element of the ensemble is probed, and as we will see, there is some large timescale at which typical elements of the ensemble give different probe measurements from the thermal average.    Of course, the {\it average} over many measurements in the same ensemble of universes will accurately reproduce the thermal expectation value.  This average will thus agree with computations made in the thermo-field formalism,  the two entangled CFTs of this formalism being dual to the eternal black hole.   Also,  {\it semiclassical} observers having access only to coarse grained observables will be unable to  measure the differences between the microstates and hence  will describe their results in terms of the eternal black hole even if this description is  microscopically incorrect.

In Sec.~2 we discuss different natural ensembles  and show how the entropic suppression of variance arises in a general theory. In Sec.~3 and Sec.~4 we use these results to extract the timescales at which correlation functions computed in  microstates of a theory differ from the thermal average and from each other.  To build intuition, Sec.~3 focuses on a simple example, the free boson on a circle.  Sec.~4 discusses the microstates of the BTZ $M=0$
black hole.\footnote{This extends results of \cite{d1d5} by
showing how the entropic suppression of variance arises.}    While
the latter computations are done in the free limit of the dual
symmetric product sigma model, the ``fractionation'' of string
bits associated to different winding sectors of this CFT
effectively models the large entropy and narrow gap structure
associated to theories that describe black holes.

%~~~~~~~~~~~~~~~~~~~~~~~~~~~~~~~~~~~~~~~~~~~~~~~
\section{Ensembles and variances of observables}
\label{variances}
%~~~~~~~~~~~~~~~~~~~~~~~~~~~~~~~~~~~~~~~~~~~~~~~

Let us suppose that a measuring device with energy resolution
$\Delta E$ measures the mass of a black hole to be $E$.    In
quantum mechanics a ``measurement'' of this kind implies that the
device registered an energy eigenvalue lying between $E$ and $E +
\Delta E$.\footnote{The ensemble of microstates associated to a black hole may not always be specified by macroscopic local conserved charges.  See \cite{susybring, babel,dipole} for a discussion within the AdS/CFT correspondence, where states are also characterized by non-conserved dipole charges.}  All pure microstates consistently giving such energy
measurements are superpositions of a basis of energy eigenstates
\begin{equation}
\CM_{bas} = \left\{ \  |s\rangle  \ : \ H |s \rangle = e_s |s \rangle ~~~;~~~ E \leq e_s \leq E + \Delta E \
\right\} \, ,
\Label{Eeigenstates}
\end{equation}
and are thus elements of the ensemble
\begin{equation}
\mcont  =
\left\{|\psi\rangle = \sum_s c^\psi_s  |s\rangle
\right\} \, ,
\Label{allstates}
\end{equation}
with $|s\rangle$ as in (\ref{Eeigenstates}) and $\sum_s |c_s|^2 =
1$.     The expectation value of the Hamiltonian $H$ in any state
in $\mcont$ also lies between $E$ and $E+\Delta E$.  If entropy of
the system\footnote{We are working in the
microcanonical ensemble of states of fixed energy (within the
resolution $\Delta E$).    It is worth mentioning that in flat
space (unlike AdS space \cite{hawkingpage}) the canonical ensemble
is ill-defined for the physics of black holes because the growth
of the density of states is too rapid to give a convergent
partition function \cite{wald}. In the context of this paper where
we will test to what extent the variance in observables can
distinguish black hole microstates, the canonical ensemble is also
awkward to use because, for some observables, the variance may be
dominated by the deviations in energy within the ensemble rather than by
deviations in structure between states of a fixed energy. We will
return to this briefly in Appendix~\ref{realisticth}.} is $S(E)$, then the basis in (\ref{Eeigenstates}) has
dimension $e^{S(E)}$:
\begin{equation}
1+ \dim{\CM_{sup}} = | \CM_{bas} | = e^{S(E)}\, .
\Label{numstates}
\end{equation}
It has been argued \cite{circularity} that generic ``foam"
microstates of the BTZ $M=0$ black hole \cite{fuzzball} correspond
to quantum superpositions of the chiral primary operators in the
dual CFT. In that case, all the superposed basis states have the
same energy, while in the ensemble (\ref{allstates}) we are also
permitting superpositions of microstates of different energies
that lie within the measurement resolution $\Delta E$. We would
like to evaluate the spread in observables, taken to be finitely
local correlation functions of Hermitian operators, measured in
the different microstates of a black hole, as well as the
differences with measurements made in a thermal state.\footnote{We collect the basic definitions of various special states and ensembles used in the text in Appendix \ref{taxonomy}.}

\subsection{The variance in quantum observables}

\subsubsection{Variances in quantum mechanics}
To begin, consider quantum mechanical measurements, made by a
Hermitian operator $\CO$, taken in an ensemble of microstates. If the
microstate $|\alpha\rangle$ is an eigenstate of $\CO$, the measurement
gives the eigenvalue $o_\alpha$, i.e.,
\begin{equation}
\CO \, | \alpha \rangle = o_\alpha \, |\alpha\rangle
\end{equation}
In general, $\CO$ and the Hamiltonian will not commute ($[\CO,H]
\neq 0$);  hence there will not be a basis of simultaneous
eigenstates of $\CO$ and $H$.    Thus, to characterize measurement
by $\CO$ we have to take a perspective wherein the universe is
repeatedly prepared in an identical microstate which is repeatedly
probed by $\CO$, leading each time to a different eigenvalue
$o_t$.    Any given microstate has an expansion
\begin{equation}
|\psi\rangle = \sum_\alpha c_\alpha^\psi \, |\alpha\rangle
\Label{psiexp}
\end{equation}
so that the probability of measuring eigenvalue $o_\alpha$ when the
underlying state is $|\psi\rangle$ is $|c_\alpha^\psi|^2$.

Over the entire ensemble of states $\mcont$, repeated measurement
gives a spectrum of measured eigenvalues.   The spread in these
eigenvalues over $\mcont$ can be characterized by the variance and
mean of the distribution of eigenvalues over the entire ensemble.
The ensemble average is
\begin{equation}
\langle \CO \rangle_{\mcont} = \sum_\alpha \, \Pr(\alpha) \,
o_\alpha = \sum_\alpha  \left( \int  D\psi \,  |c_\alpha^\psi|^2
\right) o_\alpha = \int D\psi \, \langle \psi | \CO | \psi \rangle
\, .
\end{equation}
Here $\Pr(\alpha)$ is the probability of $|\alpha\rangle$ appearing in any state
within the ensemble $\mcont$, and $\int D\psi$ indicates an
integral over all states in the ensemble with the normalization
that $\int D\psi = 1$.   The variance in eigenvalues of $\CO$ over
the ensemble $\mcont$ is likewise
\begin{eqnarray}
{\rm var}[\CO]_{\mcont} &=&
\sum_\alpha \Pr(\alpha) \, o_\alpha^2 -  \left( \sum_\alpha \Pr(\alpha) \, o_\alpha \right)^2
=
\int D\psi \, \langle \psi | \CO^2 | \psi \rangle  - \langle \CO \rangle_{\mcont}^2 \\
&=&
\langle \CO^2 \rangle_{\mcont} - \langle \CO \rangle_{\mcont}^2
\end{eqnarray}
This ensemble variance in eigenvalues of $\CO$ characterizes how
widely the ensemble $\mcont$ is  spread over eigenvectors of
$\CO$.   However, it does {\it not} characterize how different the
individual states in $\mcont$ are from each other in their
responses to being probed by $\CO$.  Hence a different notion of
variance is necessary.

Any given state in $\mcont$ responds to $\CO$ by producing the
eigenvalue $o_t$ with probability $|c_t^\psi|^2$.    Thus we
really want to characterize the differences in the probabilities
for measuring $o_t$ in the different states.    These probability
distributions are equally characterized by their moments:
\begin{equation}
c^0_\psi = \langle \psi | 1 | \psi \rangle \, , \
c^1_\psi = \langle \psi | \CO | \psi \rangle \, , \
c^2_\psi = \langle \psi | \CO^2 | \psi \rangle \, , \
c^3_\psi = \langle \psi | \CO^3 | \psi \rangle \,  , \, 
\ldots
\Label{momentdef}
\end{equation}
We would like to measure how these moments vary over the ensemble
$\mcont$.   The ensemble averages of the moments (\ref{momentdef})
and their variances over the ensemble are given by
\begin{eqnarray}
\langle c^k\rangle_{\CM_{sup}} &=& \int D\psi \, c_\psi^k \Label{momentdef2} \\
{\rm var}[c^k]_{\CM_{sup}}
&=& \int D\psi \, (c_\psi^k)^2 -
\langle c^k \rangle_{\CM_{sup}}^2 \, .
\Label{vardef2}
\end{eqnarray}
To compute these quantities, we first construct the generating function
\begin{equation}
C_\psi(\theta) = \sum_n {\theta^{\,n} \over n!} c^n_\psi = \langle
\psi | e^{\theta \CO} | \psi \rangle \, , \Label{momentgen}
\end{equation}
and its ensemble average
\begin{equation}
\langle C(\theta) \rangle_{\CM_{sup}} = \int D\psi \,
C_\psi(\theta) = \sum_t \Pr(t) \, e^{\theta\, o_t} = \sum_n
{\theta^{\,n} \over n!} \langle c^n \rangle_{\CM_{sup}} \, .
\Label{momentgenav}
\end{equation}
We can also define
\begin{equation}
\langle C_2(\theta_1,\theta_2) \rangle_{\CM_{sup}} = \int D\psi \,
C_\psi(\theta_1) \, C_\psi(\theta_2) - \langle C(\theta_1)
\rangle_{\CM_{sup}} \langle C(\theta_2) \rangle_{\CM_{sup}}
\Label{vargenav}
\end{equation}
In terms of (\ref{momentgenav}, \ref{vargenav}) the ensemble
averages (\ref{momentdef2}, \ref{vardef2}) are
\begin{eqnarray}
\langle c^k \rangle_{\CM_{sup}} &=& \left[ {d^k \langle C(\theta)
\rangle_{\CM_{sup}} \over d\theta^k}  \right]_{\theta = 0}
\Label{momentav} \\
{\rm var}[c^k]_{\CM_{sup}} &=& {d^k \over d \theta_1^k} {d^k \over
d\theta_2^k}  \left[  \langle C_2(\theta_1,\theta_2)
\rangle_{\CM_{sup}} \right]_{\theta_1=\theta_2 = 0} \, .
\Label{varav}
\end{eqnarray}
The differences between states in the ensemble of microstates in
their responses to local probes are quantified by the
standard-deviation to mean ratios
\begin{equation}
{\sigma[c^k]_{\CM_{sup}} \over \langle c^k \rangle_{\CM_{sup}}} =
{\sqrt{{\rm var}[c^k]_{\CM_{sup}}} \over \langle c^k
\rangle_{\CM_{sup}}} \, .
\Label{proxydev}
\end{equation}

\subsubsection{Variances in quantum field theory}
We would like to extend the quantum mechanical definition of
variance to the  finitely local correlation functions that are the
observables of relevance to us:
\begin{equation}
c_\psi^k(x^1,\ldots , x^k) = \langle \psi| \CO(x^1) \cdots \CO(x^k) | \psi \rangle \, .
\end{equation}
These are the field theory analogues of the moments
(\ref{momentdef}) of the eigenvalue distribution in a quantum
mechanical state. Following the previous section, the differences
in the correlation function responses of states  in $\mcont$ are
quantified by the means and variances
\begin{eqnarray}
\langle c^k(x^1,\ldots , x^k) \rangle_{\CM_{sup}} &=& \int D\psi  \, c_\psi^k(x^1,\ldots , x^k)
\Label{moment2}
\\
{\rm var}[c^k(x^1,\ldots , x^k)] _{\CM_{sup}}  &=&
 \int D\psi \,(c_\psi^k(x^1,\ldots , x^k))^2 -
 \langle c^k(x^1,\ldots , x^k) \rangle_{\CM_{sup}} ^2
 \Label{var2}
\end{eqnarray}
The only difference from the quantum mechanical case is that the
observables are now functions of spacetime coordinates.   As
before, we imagine preparing the universe repeatedly in a black
hole microstate $|\psi\rangle$ and making repeated measurements
with a local operator $\CO$ to probe the state.  These quantities
(\ref{moment2}, \ref{var2}) are written more efficiently in terms
of the generating function
\begin{equation}
Z_\psi[J] = \langle \psi| e^{\int dx \, J(x)  \CO(x)} | \psi \rangle = \langle \psi | \, Z[J] \, | \psi
\rangle
\end{equation}
and the associated ensemble averages
\begin{eqnarray}
\langle Z[J]\rangle_{\CM_{sup}} &=& \int D\psi \, Z_\psi[J]
\Label{momentav2}
\\
\langle Z_2[J_1,J_2]\rangle_{\CM_{sup}} &=& \int D\psi \,
Z_\psi[J_1]  Z_\psi[J_2] - \langle Z[J_1]\rangle_{\CM_{sup}} \,
\langle Z[J_2]\rangle_{\CM_{sup}} \Label{varav2}
\end{eqnarray}
Following (\ref{momentav}, \ref{varav}), appropriate functional
derivatives of (\ref{momentav2}, \ref{varav2}) give
(\ref{moment2}, \ref{var2}).  Below we will show that the standard
deviation to mean ratio
\begin{equation}
{\sigma[c^k(x^1,\ldots , x^k)]_{\CM_{sup}} \over \langle c^k(x^1,\ldots , x^k) \rangle_{\CM_{sup}}}
=
{\sqrt{{\rm var}[c^k(x^1,\ldots , x^k)]_{\CM_{sup}}} \over \langle c^k(x^1,\ldots , x^k) \rangle_{\CM_{sup}}}
\end{equation}
is heavily suppressed because generic states are statistically
random quantum superpositions and hence lie very close to a
certain ``typical'' state.

\subsection{Entropic suppression of variance}

The integrals $D\psi$ in the generating functions of the means and
variances of the moments $c^k$ (\ref{momentav2}, \ref{varav2}) can
be evaluated in the energy basis (\ref{Eeigenstates}) by
integrating over the superposition coefficients $c_s^\psi$ in
(\ref{allstates}).  This gives
\begin{eqnarray}
\langle Z[J]\rangle_{\CM_{sup}} & = &
  \int d\vec{c}^{\, \psi} \,
     \sum_{s\, t} c_s^\psi c_t^{\psi\, *}
       \langle t |
       e^{\int J  \mathcal{O}}
        | s \rangle
          \Label{meanav3}
               \\
\langle Z_2 [J_1,J_2]\rangle_{\CM_{sup}} & = &
  \int\, d\vec{c}^{\, \psi} \, \sum_{s\, t\, m\, n}
     c_s^\psi c_t^{\psi\, *} c_m^\psi c_n^{\psi\, *} \,
       \langle t | e^{\int J_1\mathcal{O}} | s \rangle \,
       \langle n | e^{\int J_2\mathcal{O}} | m \rangle \nonumber \\
& & - \, \langle Z[J_1]\rangle_{\CM_{sup}} \,
         \langle Z[J_2]\rangle_{\CM_{sup}}
      \Label{varav3}
\end{eqnarray}
Here  $\sum_s |c_s^\psi|^2 = 1$ and the measure is normalized to
$\int d\vec{c}^{\,\psi} = 1$. With these conventions,
\begin{eqnarray}
\int\, d\vec{c}^{\,\psi} |c_s^\psi|^2 &=& \frac{1}{e^S} \nonumber \, , \\
 \int\, d\vec{c}^{\,\psi} |c_s^\psi|^2 \, |c_t^\psi|^2 &=&
       \frac{1 + \delta_{st}}{e^S(e^S+1)}\,  . 
\end{eqnarray}
Many terms in (\ref{meanav3}, \ref{varav3}) vanish due to
integrations over the phases of $c_s^\psi$, leaving
\begin{eqnarray}
\langle Z[J]\rangle_{\CM_{sup}} & = &
  \frac{1}{e^S} \sum_s
      \langle s | e^{\int J \mathcal{O}} | s \rangle
            \Label{meanav5}
\\
      \langle Z_2 [J_1,J_2]\rangle_{\CM_{sup}} &=&  {1 \over e^S + 1} \,
\langle Z_2 [J_1,J_2]\rangle_{\CM_{bas}}\, ,
\Label{varav5}
\end{eqnarray}
where
\begin{eqnarray}
\langle Z_2 [J_1,J_2]\rangle_{\CM_{bas}} = & & { 1 \over e^{S}}
\sum_s \langle s | e^{\int J_1 \mathcal{O}} | s\rangle \langle s |
e^{\int J_2 \mathcal{O}} | s\rangle - {1 \over e^{2S}} \sum_{s\neq
t} \langle s | e^{\int J_1 \mathcal{O}} | s\rangle
\langle t | e^{\int J_2 \mathcal{O}} | t\rangle \nonumber \\
& + & {1 \over e^S} \sum_s \langle s | e^{\int J_1 \mathcal{O}} \,
\mathbb{P}_E^s \, e^{\int J_2 \mathcal{O}} | s \rangle \, .
\Label{varav6}
\end{eqnarray}
Here $\mathbb{P}_E^s = \sum_{t} | t \rangle \langle t |   -
|s\rangle\langle s|$ is a projector onto the subspace of ${\cal
M}_{sup}$ that is orthogonal to $|s\rangle$.    Taking the
correlation functions in each term in (\ref{varav6}) to be of
$\mathcal{O}(1)$,  $\langle Z_2 [J_1,J_2]\rangle_{{\cal M}_{bas}}
$ is also a quantity of $\mathcal{O}(1)$.   Then $\langle
Z_2[J_1,J_2] \rangle_{{\cal M}_{sup}}$  is exponentially
suppressed by a factor of $e^{-S}$.      This entropic suppression
is inherited by the variances of all correlation functions.

The first line in (\ref{varav6}) is precisely the generating
function of the variance of correlation functions in the ensemble
of basis elements ${\cal M}_{bas}$.   The second line can be taken
to vanish if  $[\mathcal{O},H] = 0$ because we could pick
$|s\rangle$ to be a basis of joint eigenstates of $\mathcal{O}$
and $H$.     When $[\mathcal{O},H] \neq 0$, the second line in
(\ref{varav6}) leads to a contribution to the variance of any
$k$-point correlator that is of the general form
\begin{equation}
 {1 \over e^S}
\sum_{s\neq t} \langle s | c^k  |t \rangle \langle t | c^k |s \rangle
\leq
 {1 \over e^S}
  \sum_{s}  \langle s | c^k   c^k |s \rangle  \, .
\end{equation}
The inequality follows by recognizing that each term in the sum is
positive definite and by extending the sum over $t \in \CM_{bas}$
to a complete sum over all states in the Hilbert space.    Thus we
recognize that the only avenue to having a variance large enough
to distinguish microstates by defeating the $e^S$ suppression in
(\ref{varav5}) is to find probe operators that have exponentially
large correlation functions.

\paragraph{Summary: }  Given the macroscopic quantum numbers of a system (with some measurement resolution) the generic microstate can be written as a random superposition of some basis of states with eigenvalues in the measured range.    We have shown on general grounds that the variance of local correlators in the ensemble of all microstates is suppressed relative to the variance in the ensemble of basis elements by a factor of $e^{-S}$.      In Appendix A we demonstrate the conditions under which  this conclusion continues to hold even in ensembles where only the expectation value (as opposed to the actual eigenvalues) of macroscopic observables are fixed.    For black holes with their enormous entropy, this means that unless a probe correlation function is intrinsically exponentially large, a semiclassical observer will have no hope of telling microstates apart from each other.\footnote{Of course a high energy observer with access to very high resolution will be able to separate the tiny differences between microstates.}    Correlation functions in real time typically do not grow in this way and hence will not provide suitable semiclassical probes.  However,  correlations can grow exponentially with imaginary time.   Hence, and motivated by previous attempts to probe the singularities of black holes \cite{fhks, festucciaL, mosheetc, maeda, leviross}, in the sections below we will explore whether the imaginary time behavior of correlation functions will more readily separate the microstates from each other and from the thermal ensemble average.

%~~~~~~~~~~~~~~~~~~~~~~~~~~~~~~~~~~~~~~~~~~~~~~
\section{Free scalar toy model}
\label{freescalar}
%~~~~~~~~~~~~~~~~~~~~~~~~~~~~~~~~~~~~~~~~~~~~~~

The entropic suppression of variance described above can be
illustrated in a simple toy model: the free chiral boson on a
circle of circumference $2\pi$.  This theory, being free, has a
highly degenerate spectrum with $O(1)$ gaps.   We will pick the
energy resolution of the ensemble (\ref{Eeigenstates}) so as to
focus on the degenerate states of a fixed energy $E$.     The
Lagrangian is
\eqn{boslag}{
\CL = \int  \, dt \, dx \, \p_{\mu} \vp \, \p^{\mu} \vp \ . }
The mode expansion for right-movers reads:
\eqn{modeexp}{ \vp(t,x)  =  \frac{a_0}{2} +  \sum_{n=1}^\infty \,
\frac{1}{\sqrt{n}} \, \(a_n \, e^{in\,\sigma} + a_n^\dagger \, e
^{-in\, \sigma}\)}
with $\sigma=x-t$, and the canonical commutation relations are
\eqn{commrel}{ [a_m, a_n^\dagger ] = \d_{mn} \ , \ \;\;\;\;
[a_m,a_n] = [a_n^\dagger, a_m^\dagger] = 0 \ .}

%-----------------------------------------------------------
\subsection{Correlation functions in a typical state}
\label{typcorr}
%-----------------------------------------------------------

We investigate the correlation functions of local time-ordered
right-moving operators $\CO$ built out of $\vp(\sigma)$ and its
derivatives. The states
\eqn{purefree}{ \ket{s} = \prod_{n=1}^\infty
\frac{(a_n^\dagger)^{N_n}}{\sqrt{N_n!}} \ket{0} \, , }
normalized to $\langle s | s \rangle = 1$ and subject to the
microcanonical constraint $\sum_{n=1}^{\infty} \, n \, N_n = E$,
span the $E$-eigenspace of the Hamiltonian and form the ensemble
$\mdiscr$.   For simplicity we concentrate on the primary operator
$\p \vp$. The two point function at zero spatial separation is
\eqn{primcorr}{ \mathcal{O}(t) = \langle s |\, T \[\p \vp^\dagger
(t) \, \p\vp(0) \] | s \rangle =
 \sum_{n=1}^\infty \, n \, \( e^{-i n t} + 2 \, N_n \,\cos{nt}
 \) .}
(The spatial periodicity translates into  $2\pi$-periodicity in
lightcone coordinates.)   The correlation functions \primcorr\
are linear combinations of the occupancies $N_n$.   We will be
interested in comparing correlation functions in a generic
microstate with the result for a thermal ensemble.    While the
elements of $\mdiscr$ are characterized by integer values of
$N_n$, we can define a ``thermal state'' that is obtained by the substitution:
\eqn{typicalN}{  N_n  \rightarrow  \langle N_n \rangle = \kappa(n)
=\frac{1}{e^{\b \, n} -1} \;\;\; \longmapsto \;\;\;\;
\sum_{n=1}^{\infty} \, n \, N_n = E}
This is a formal manipulation because the occupation numbers of a
specific microstate must be integers, but the expectation values
in \typicalN\ are not so constrained.   The ``thermal state'' is useful because it will reproduce the behaviour of expectation values taken in the thermal ensemble.  A more precise statistical derivation of the ``thermal state'' is given in Appendix ~\ref{typmicro} where it is shown that 
\begin{equation}
\beta = \sqrt{{\zeta(2) \over E}}
\end{equation}
and that the entropy associated to the states of energy $E$ is
\begin{equation}
S = {2 \zeta(2) \over \beta} + O(\log\beta)
\end{equation}
We are also interested in the behavior of the correlation function in the complex time plane since this might help us to defeat the entropic suppression of
variances demonstrated in Sec.~2. Continuing to imaginary time $t
\rightarrow -i \tau$, at zero spatial separation ($x = x'$) we get
\eqn{typcorrim}{ \mathcal{O}(\tau) = \langle s | \, T \[
\p\vp^\dagger (\tau) \, \p\vp(0)\] | s \rangle =
 \sum_{n=1}^\infty \, n \, \( e^{- n \, \tau } + 2 \, N_n \,\cosh  n \,\tau  \) \ .
}

\paragraph{Thermal state correlators: }  As a check, we can compare the correlator
\typcorrim\ subject to \typicalN, with the thermal answer in
two-dimensional field theory.   The thermal propagator $\Cther^\vp (\tau , x)=
\expval{\vp(0,0) \, \vp(\tau,x)}$ in momentum space is
\eqn{thermalprop}{ \Cther^\vp\(i\, \om_n, p\) = \frac{1}{\om_n^2 +
p^2} }
with $\om_n = \frac{2\pi\,n}{\beta}$, the Matsubara frequency.
Fourier transforming, we get:
\eqn{mixedther}{\eqalign{
\Cther^\vp\(\tau,p\) &= \sum_n \, e^{-i \om_n \,\tau} \, \Cther^\vp\(i \om_n , p\) \cr
 & = \frac{1}{2 |p|} \, \[e^{- |p| \, \tau} + 2 \, \kappa(p) \cosh{p \tau}\] \cr
{\rm with} \;\;\; \kappa(p) &= \frac{1}{e^{\beta \, |p|} -1} }}
The correlation function
of the primary operator $\p \vp$ is found by taking two
$\tau$-derivatives:
\eqn{primther}{ \Cther\(\tau,p\) =  \expval{\p \vp(0,0) \, \p
\vp(\tau, p)} =  \frac{|p|}{2} \, \[e^{- |p| \, \tau} + 2 \,
\kappa(p) \cosh{p \tau}\] }
This reproduces the \typcorrim\ after recognizing that $p$ takes
discrete values and that the position space correlator involves a
Fourier transform with respect to $p$.   The sum over $p$ can be
carried out explicitly, giving the result for  zero spatial
separation:
\eqn{thzeropos}{
\Cther\(\tau,0\) = \sum_p \, \Cther\(\tau,p\) = {\pi^2\over \b^2} \,  \csc^2 {\pi \,\tau \over \b}}
%

%----------------------------------------------------------------------------------
\subsection{Variance in the microcanonical ensemble}
%----------------------------------------------------------------------------------

We would like to calculate the variance in the correlation functions in the microcanonical ensemble. As we have seen in Sec.~2, it suffices to compute the variance in $\mdiscr$ since the variance in $\mcont$ can then be obtained using the exponential suppression factor. 

\paragraph{Variance in $\mdiscr$:} From the expression for the correlation function \primcorr, we can write down the variance in the basis ensemble $\mdiscr$:
\begin{eqnarray}
 {\rm
var}(\mathcal{O}(\tau))_{\CM_{bas}} &=& \sum_{n=1}^E
\, (2 \,n \cosh{n \tau})^2 \, {\rm var}(N_n)_{\CM_{bas}} 
\nonumber \\
& &   + \, 2 \, 
\sum_{m<n} \, 4 \, n \, m \, \cosh{n \tau} \, \cosh{m
\tau} \, {\rm cov}(N_n, N_m)_{\CM_{bas}} 
\Label{variance1}
\end{eqnarray}
The covariance ${\rm cov}(N_n,N_m)$ is
\eqn{covariance}{ {\rm cov}(N_n, N_m) =\expval{N_n \, N_m} -
\expval{N_n} \, \expval{N_m} \ .}
Because of the constraint $\sum_n n N_n = E$ in the microcanonical ensemble, ${\rm cov}(N_n,N_m) \leq 0$.

To estimate (\ref{variance1}), it is useful to first evaluate the variances in the occupation numbers in the canonical ensemble since the exact distribution function is known. The standard result gives:
\eqn{covarcano}{ {\rm cov}(N_n, N_m)_{can} = \frac{\delta_{n m}
}{4\sinh^2\frac{\beta n}{2}} \, ,}
One can use these standard results from the canonical ensemble to estimate the microcanoncial variance in $\mdiscr$. To this end  write the microcanonical variance as 
\begin{equation}
{\rm var}(\mathcal{O}(\tau))_{\CM_{bas}} =  \sum_{n=1}^E \( \frac{n \cosh{n \tau}}{\sinh\frac{\beta n} {2}} \)^2 \, g(n) \, F(n) \ ,
\Label{variance2}
\end{equation}
where we define the new functions $g(n)$ and $F(n)$ as 
\begin{eqnarray}
g(n)& =&\frac{{\rm var}(N_n)_{\CM_{bas}}}{{\rm var}(N_n) _{can}}\\
F(n) & = & 1 + 2 \sum_{m < n} \frac{m \, \cosh{(m \tau)} \, {\rm
cov}(N_n, N_m)_{\CM_{bas}}}{n \, \cosh{(n \tau)} \, {\rm
var}(N_n)_{\CM_{bas}}} \ .
\Label{variance2a}
\end{eqnarray}
The point of this exercise has been to rewrite the microcanonical variances in terms of the canonical variances, in order to extract bounds on ${\rm var}(\mathcal{O}(\tau))_{\CM_{bas}}$. 

To derive an upper bound on ${\rm var}(\mathcal{O}(\tau))_{\CM_{bas}}$, note  that $g(n) < 1 $ in general (essentially because the canonical ensemble incorporates more outlying states with energies greater than $E$ making the variations more spread out) and that $F(n) \leq 1$ because  ${\rm cov}(N_n,N_m) \leq 0$.    Hence   
$
{\rm var}(\mathcal{O}(\tau))_{\CM_{bas}} <
\sum_{n=1}^E \( n \cosh(n \tau) / \sinh(\beta n/2) 
\)^2 $.   For a lower bound, first note that on general grounds the canonical ensemble will  approximate $\mdiscr$ well for small occupation numbers.   Hence $g(n) \approx 1$ and ${\rm cov}(N_n,N_m) \propto \delta_{nm}$ for $n,m$ below some threshold value $n_c$.     Then if all the terms in (\ref{variance2}) are positive, or equivalently, if  $F(n)>0 \ \forall n$, the lower bound 
$
 \sum_{n=1}^{n_c} \( n \cosh(n \tau) / \sinh(\beta n/
2) \)^2 < {\rm var}(\mathcal{O}(\tau))_{\CM_{bas}}
$ follows.

To see that $F(n)>0$, consider an auxiliary operator $\tilde{\mathcal{O}}=\sum_n N_n$. 
In a manner similar to Eqs.~(\ref{variance2}-\ref{variance2a}), the variance in 
$\tilde{\mathcal{O}}=\sum_n N_n$ at a fixed energy $E$ may be re-written in the form:
\begin{eqnarray}
{\rm var}(\tilde{\mathcal{O}})_{\CM_{bas}} & = & \sum_{n=1}^E \( \sinh\frac{\beta n} {2} \)^{-2} \, g(n) \, \tilde{F}(n) \\
\tilde{F}(n) & = & 1 + 2 \sum_{m < n} \frac{{\rm cov}(N_n, N_m)_{\CM_{bas}}}{{\rm var}(N_n)_{\CM_{bas}}} \, .
\end{eqnarray}
On the other hand, $\tilde{\mathcal{O}}$ counts the total number of excited oscillators in a given state.
Hence, its variance is an increasing function of $E$.  It follows that:
\begin{equation}
0 < {\rm var}(\tilde{\mathcal{O}})_{\CM_{bas}}|_{E+1} - {\rm var}(\tilde{\mathcal{O}})_{\CM_{bas}}|_E = 
\( \sinh\frac{\beta (E+1)} {2} \)^{-2} \, g(E+1) \, \tilde{F}(E+1) \quad \forall E \, ,
\end{equation}
so $\tilde{F}(n)>0$. A quick inspection of the definitions of $F,\, \tilde{F}$  reveals that $F(n) > \tilde{F}(n)$ so that 
$F(n)>0$ and the lower bound on ${\rm var}(\mathcal{O}(\tau))_{\CM_{bas}}$ follows.
In summary, we find 
\eqn{variancebounds} {
 \sum_{n=1}^{n_c} \( \frac{n \cosh{n \tau}}{\sinh\frac{\beta n}
{2}} \)^2 < {\rm var}(\mathcal{O}(\tau))_{\CM_{bas}} <
\sum_{n=1}^E \( \frac{n \cosh{n \tau}}{\sinh\frac{\beta n}{2}}
\)^2 \, . }

To obtain the relative magnitude of the deviations in
$\mathcal{O}(\tau)$, we divide the standard deviation by the mean
correlator as in (\ref{proxydev}). For $\tau$ not too close to 0
or $\beta$, the canonical ensemble provides a good estimate of the
latter:
\eqn{meancorrelator} { \langle \mathcal{O}(\tau)
\rangle_{\CM_{bas}} \, \approx \, \int dn\, \frac{2n \cosh{n \tau}
}{\exp{(\beta n)} -1} \, .}
Using the lower bound in \variancebounds\ we see that
$\sqrt{\rm{var}(\mathcal{O}(\tau))_{\CM_{bas}}}$ begins to grow
rapidly at $\tau = \beta/ 2$ while $\langle
\mathcal{O}(\tau) \rangle_{\CM_{bas}}$ does not undergo such
growth until $\tau > \beta$. Thus, for $\tau > \beta / 2$,
\begin{equation}
\sqrt{\rm{var}(\mathcal{O}(\tau))_{\CM_{bas}}} / \langle
\mathcal{O}(\tau) \rangle_{\CM_{bas}} \gg 1
\end{equation}
and the probe can distinguish generic members of $\mdiscr$ from
the thermal mean and from one another.

\paragraph{Variance in $\mcont$:} According to (\ref{varav5}), the variance in the ensemble of all states $\mcont$ will be suppressed by a factor of $e^S = e^{2 \beta E}$ relative to the
variance in $\mdiscr$, where we have used $S = 2\beta E$ as shown in Appendix~B. Using equation \variancebounds , this leads
to:
\eqn{maxvarcont}{(2\tau - \beta) n_c - 2 \beta E\, < \, \log\, {\rm
var}(\mathcal{O}(\tau))_{\CM_{sup}}\, <\, (2\tau - 3\beta)E\, .}
To insure that the variance not be exponentially suppressed, one
must wait at least until $\tau = {3 \beta / 2}$. However, by that
time the mean correlator will also have grown exponentially large making separation of states difficult.  In particular the late time growth of the correlation function is largely determined by the highest energy oscillator level that is populated.  To extract any further information about the state beyond that, despite the relatively large variance, higher precision measurements will be needed.

\paragraph{Summary for the free chiral boson:} In the free chiral boson theory in 1+1 dimensions we can generate a large degeneracy of states by working at energies $E \gg 1$. 
Given this degeneracy we would like to ask if it is possible to distinguish the states that make up the microcanonical ensemble at energy $E$ from each other and from the thermal state with mean energy $E$. There are two noteworthy points in our discussion: 
\begin{itemize}
\item Since the thermal correlation functions are required to be periodic in imaginary time, they diverge at $\tau = \b$ (for zero spatial separation). This behaviour is absent in each one of the microcanonical states at energy $E$.
\item The variances in the microcanonical ensemble of all superpositions ($\mcont$) start to grow exponentially at time scales of order $\tau \sim {3\b \over 2}$.   At this point we might be able to resolve the highest energy oscillator that is excited in the microstate. Further distinctions would require making either more measurements on the system or a finer resolution scale of the measuring apparatus.
\end{itemize} 
We derived these results for a particular simple observable in the free boson theory, but we expect that they will also apply to other finitely local correlators.  

%----------------------------------------------------------------------------------
\section{D1-D5 system and the BTZ $M=0$ black hole}
\label{d1d5sys}
%----------------------------------------------------------------------------------

We now study the variance of correlation functions measured in microstates associated to the BTZ $M=0$ black hole.   We will calculate this variance in the dual field theory, i.e., the D1-D5 system (reviewed e.g. in  \cite{d1d5,luninmathurcorrs}).

\subsection{Review}

Type IIB string theory on $\ads{3} \times S^3 \times T^4$ is dual to the D1-D5 CFT, a marginal deformation of the (1+1)-dimensional orbifold sigma model with target space
\begin{equation}
\mathcal{M}_0 = (\textrm{T}^4)^N / \textrm{S}_N,
\end{equation}
Here $\textrm{S}_N$ is the symmetric group of $N$ elements and $N$ is related to the AdS scale.    When this duality arises from a decoupling limit of the theory of $N_1$ D1-branes and $N_5$ D5-branes wrapped on the $S^1 \times T^4$ factor of $R_t \times R^4 \times S^1 \times T^4$, we also have $N = N_1 N_5$.     At the orbifold point the D1-D5 CFT describes string theory in a highly curved AdS space.  A supergravity description of the latter is only strictly valid at low curvature, when the marginal deformation of the dual has been turned on.   We will be calculating correlation functions and their variances at the orbifold point where the theory is free.  Hence, exact agreement with supergravity computations is expected only for quantities that are BPS protected.

The microstates of the BTZ $M=0$ black hole embedded in $\ads{3} \times S^3 \times T^4$ are dual to the ground states of the D1-D5 CFT in the Ramond sector.    In the orbifold limit, these states are constructed in terms of a set of 
bosonic and fermionic twist operators $\{\sigma_n^{\mu}, \,
\tau_n^{\mu} \}$.  A general ground state is given by
\begin{equation}
\label{ground}
\sigma(N_{n\mu},N_{n\mu}') = \prod_{\mu,n} (\sigma_n^{\mu})^{N_{n\mu}} (\tau_n^{\mu})^{N_{n\mu}'}.
\end{equation}
In the above the subscript $n \le N$  and $\sigma_n$ ($\tau_n$) cyclically permutes $n\leq N$ copies of the  CFT on a single $T^4$.  The superscript $\mu = 1,\ldots,8$ labels the
possible polarizations of the operators. Therefore a ground state
in the Ramond sector is uniquely specified by the numbers
$N_{n\mu}$ and $N_{n\mu}'$, which must be such that the total
twist equals $N$:
\begin{equation}
\sum_{n,\mu} n(N_{n\mu} + N_{n\mu}') = N, \quad N_{n\mu} = 0,1,2,\ldots, \quad N_{n\mu}' = 0,1.
\end{equation}

We are interested in the case when the total twist $N$ is large,
which translates to an exponentially large number of Ramond ground
states:   $e^{S} = e^{2\pi\sqrt{2} \sqrt{N}}$. This enables us to treat the
system statistically, following \cite{d1d5}. While we
would prefer to work in the microcanonical ensemble with only states
of total twist $N$, it is much easier to consider a canonical
ensemble of states of all possible twists. We will see later that
``fractionation'' of the CFT simplifies matters for us in this case: due to the
fractional frequencies the canonical variance of the correlator
(\ref{gravitoncorr}) remains under control for  any finite value of
imaginary time $\tau$ unlike in the free boson case.\footnote{In the previous section we focussed on estimating results in the microcanonical ensemble for this reason.   The canonical correlation functions for the free boson can actually give divergent variances at imaginary time $\beta/2$.   These divergences arise in this case from the artifact that the canonical ensemble includes states with unbounded energies, albeit suppressed exponentially in the ensemble.}   This enables
us to perform the computation using the canonical ensemble, taking
the entropic suppression of (\ref{varav5}) into account at the
end.

In the canonical ensemble the total twist is fixed by introducing
a Lagrange multiplier $\beta$, the inverse `temperature', and it
was shown in \cite{d1d5} that this relates $N$ and $\beta$ as
follows:
\begin{equation}
N = - \frac{\partial}{\partial \beta} \ln Z(\beta) \approx \frac{2\pi^2}{\beta^2}, \quad \textrm{for } \beta \ll 1.
\end{equation}
Note that $1/\beta$ is {\it not} a physical temperature.   We are dealing with the zero-temperature, massless BTZ black hole, and $\beta$ here is simply a parameter introduced to fix the total twist $N$ using the ``trick" of the canonical ensemble.  Since the twist operators are independent, their average
distributions are given simply by Bose--Einstein and Fermi--Dirac
distributions:
\begin{equation}
N_{n\mu} = \frac{1}{e^{\beta n} - 1}, \quad N_{n\mu}^\prime = \frac{1}{e^{\beta n} + 1}.
\Label{BEFD}
\end{equation}
Note that this situation is completely analogous to the one we
encountered in section 3 with the free scalar field.   Here the
excitation numbers are replaced by the twist numbers, which are
integers for any given state of the form (\ref{ground}), but
non-integral for the ``average'' state (\ref{BEFD}). Thus the reasoning of section
2 applies also to this system.

\begin{figure}[htb]
\begin{center}
 \begin{quote}
\begin{tabular}{c@{~~~~~~~~}c}
  \epsfxsize=7cm \epsfbox{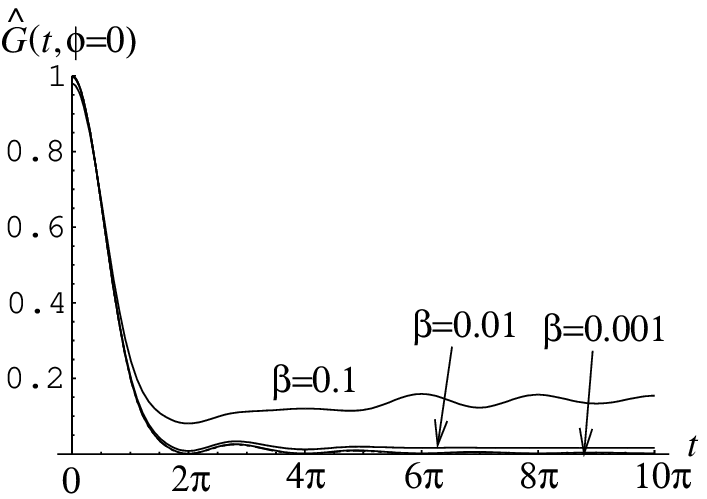}
 &
  \epsfxsize=7cm \epsfbox{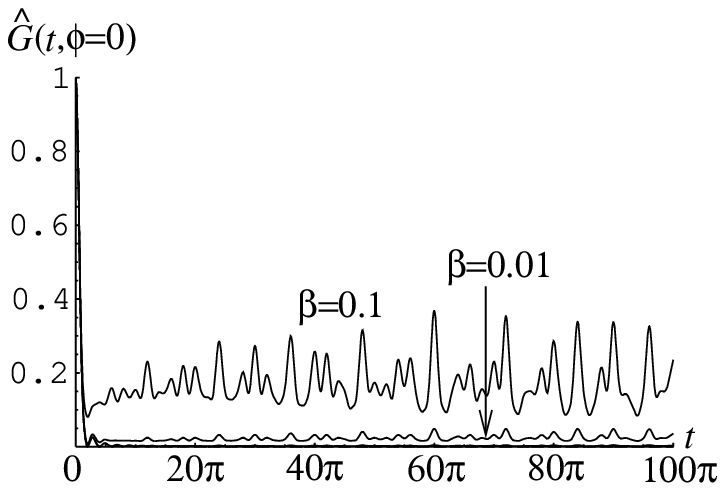}
     \\
  \epsfxsize=7cm \epsfbox{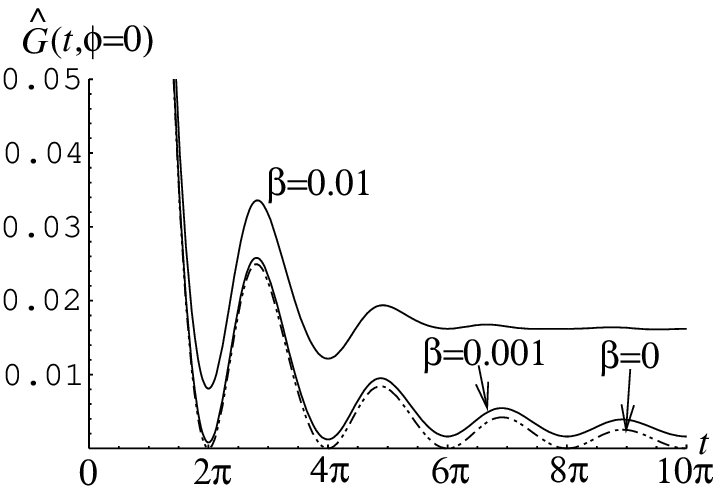}
 &
  \epsfxsize=7cm \epsfbox{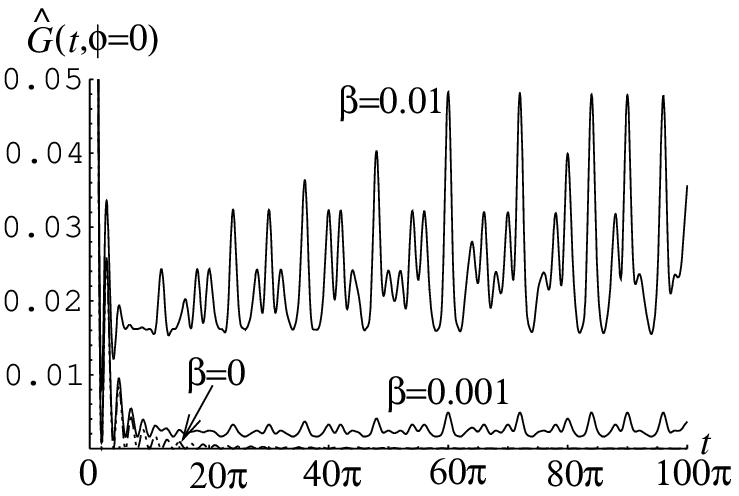}
\end{tabular}
 \caption{\sl The normalized correlation function $
\hat{G}_{(N_{n\mu},N_{n\mu}')}(t,\phi=0;\beta)$ as a function of
time $t$ for various values of the temperature
 $\beta$. The graph is reproduced from
 \cite{d1d5} with kind permission of the authors.}
 \label{Ghat}
  \end{quote}
\end{center}
\end{figure}

The key result we borrow from \cite{d1d5} is the correlation
function for a probe graviton operator $\mathcal{A} =
\partial X^a_A(z) \bar{\partial} X^b_A(\bar{z})$ in a state
specified by twist numbers $\{ N_{n\mu}, N_{n\mu}'\}$. The correlation function was computed
to be
\begin{eqnarray}
& & \hat{G}_{(N_{n\mu},N_{n\mu}')}(t,\phi) =  \langle \{ N_{n\mu},
N_{n\mu}' \} |  \mathcal{A}^{\dag}(t,\phi) \mathcal{A}(0,0) | \{
N_{n\mu}, N_{n\mu}' \}
\rangle \nonumber \\
& & = \frac{1}{N} \sum_{n=1}^{\infty} \frac{nN_n}{\left( n \sin
\frac{t}{n} \right)^2} \left[ \sin^2 \frac{\phi-t}{2} + \sin^2
\frac{\phi + t}{2} - \frac{2 \sin t \sin \frac{\phi-t}{2} \sin
\frac{\phi+t}{2}}{n \tan \frac{t}{n}} \right],
\label{gravitoncorr}
\end{eqnarray}
where $N_n = \sum_{\mu} (N_{n\mu} + N_{n\mu}')$ and the
normalization was chosen so that for the vacuum $\hat{G}_{N_n=0} =
1$. From now on we shall set $\phi=0$ for simplicity. This
correlator, evaluated for the typical state (\ref{BEFD}), is plotted in figure \ref{Ghat}, and for short time
scales of order $t \lesssim \mathcal{O}(\sqrt{N})$ the behavior of
the correlator is the universal decay expected from the BTZ black
hole.   For larger times the correlation function behaves in
a quasi-periodic way, differing from the expected behavior of the
black hole.

The main difference between this system and the free boson is that
the D1--D5 system has a much greater degeneracy of states at a
given energy scale. This is due to fractionation: whereas for the
free boson the energies come naturally in units of $\frac{1}{R}$,
for the D1--D5 system the unit size is $\frac{1}{N_5 R}$, where
$R$ is the size of the $S^1$ factor. The reason for this
is most easily understood by performing a U-duality on the D1--D5
system, which results in an FP system in type II, with a
fundamental string wound $N_5$ times the $S^1$ and $N_1$ units of
momentum going around the string. Since the total length of the
string is $N_5 R$, the energies associated to this system are
naturally quantized in units of $\frac{1}{N_5
R}$.   Mathematically this translates to having
fractional frequencies in the correlator (\ref{gravitoncorr}),
whereas for the free boson the frequencies were integral
(\ref{primcorr}).

\subsection{Euclidean variance}
We now wish to analyze the variance and the standard deviation to
mean ratio of the correlator (\ref{gravitoncorr}) in both real
and imaginary time. We start by rotating to Euclidean time by $t \to -i \tau$. Using the well known
results for the canonical variances of the Bose--Einstein and Fermi--Dirac
distributions,
\begin{equation}
\textrm{var}(N_{n\mu})_{can} = \frac{1}{4 \sinh^2 \frac{\beta n}{2}},
\quad \textrm{var}(N_{n\mu}')_{can} = \frac{1}{4 \cosh^2 \frac{\beta
n}{2}}, \label{varN}
\end{equation}
we can compute the variance of the Euclidean correlator. It is
given by
\begin{equation}
\textrm{var}(\hat{G}_{(N_{n\mu},N_{n\mu}')}(\tau))_{can} = \frac{64
\beta^4}{\pi^4} \sinh^4 \frac{\tau}{2}  \sum_{n=1}^{\infty}
\frac{1}{n^2 \sinh^4 \frac{\tau}{n}} \left( 1+ \frac{\sinh \tau}{n
\tanh \frac{\tau}{n}} \right)^2 \frac{\cosh \beta n}{\sinh^2 \beta
n}. \label{gravitonvariance}
\end{equation}
\begin{figure}
\begin{center}
\includegraphics[scale=0.60]{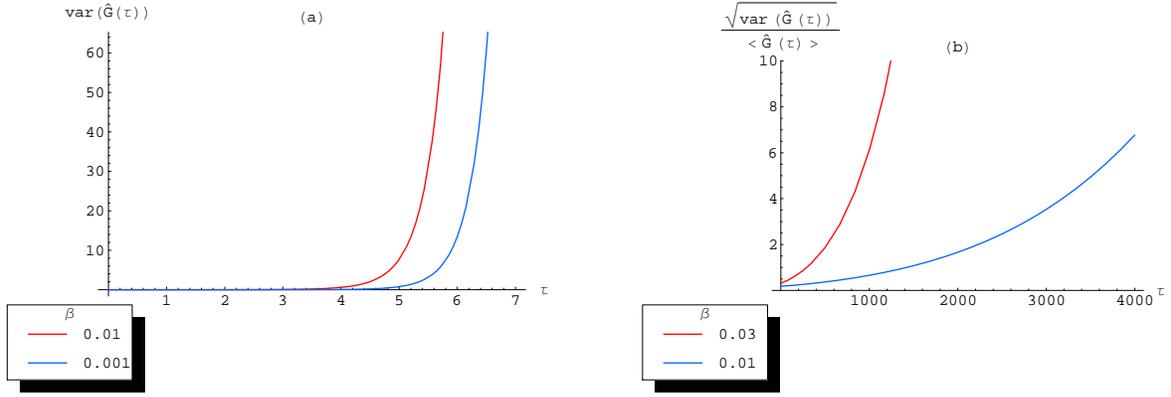}
\end{center}
\caption{\small{(a): Variance as a function of Euclidean time. The
`temperatures' are $\frac{1}{\beta} = 100$ (left) and $1000$
(right).  (b): Standard deviation to mean--ratio as a function
of Euclidean time for `temperatures' $\frac{1}{\beta} = \frac{100}{3}$
(upper) and $100$ (lower).}}
\label{eucl}
\end{figure}
As indicated by figure \ref{eucl}, this variance is exponentially growing
and the time $\tau_0$ when it begins to grow rapidly is much
smaller than $\frac{1}{\beta}$. Analyzing the sum in this regime,
$1 \ll \tau \ll \frac{1}{\beta}$, one can see that it is dominated
by terms with  $n \sim \mathcal{O}(\frac{1}{\beta})$, for which
the terms behave like $ \frac{1}{64} \beta^2
\frac{e^{4\tau}}{\tau^6}$. Since there are
$\mathcal{O}(\frac{1}{\beta})$ of these we see that the variance
approximately behaves as
\begin{equation}
\textrm{var}(\hat{G}_{N_{n\mu},N_{n\mu}'}(\tau))_{can} \sim \frac{1}{\pi^4}
\beta \frac{e^{4\tau}}{\tau^6}.
\end{equation}
From this we can conclude that the relevant timescale for rapid
growth of the variance is given approximately as $\frac{1}{4} \ln
\frac{1}{\beta} \sim \ln N^{\frac{1}{8}} \sim \ln
S^{\frac{1}{4}}$.

However, the quantity that measures the size of fluctuations in
the ensemble is the standard deviation to mean-ratio. Using
results from above we can easily compute this to be
\begin{equation}
\frac{\sqrt{\textrm{var}(\hat{G}_E(\tau))_{can}}}{\langle
\hat{G}_E(\tau) \rangle} = \frac{ \sqrt{ \sum_{n=1}^{\infty}
\frac{1}{n^2 \sinh^4 \frac{\tau}{n}} \left( 1+ \frac{\sinh \tau}{n
\tanh \frac{\tau}{n}} \right)^2 \frac{\cosh \beta n}{\sinh^2 \beta
n}}}{\sum_n^{\infty} \frac{1}{n \sinh^2 \frac{\tau}{n}} \left( 1+
\frac{\sinh \tau}{n \tanh \frac{\tau}{n}} \right) \frac{1}{\sinh
\beta n}}.  \label{stdmean}
\end{equation}
This is also plotted in figure \ref{eucl} and also exhibits exponential
growth. However, the relevant timescales are now much longer:
$\tau \gg \frac{1}{\beta}$. Analyzing the sums in this regime we
find that they are dominated by terms with $n \sim
\mathcal{O}(\tau)$, which contribute as
\begin{equation}
\frac{\sqrt{\textrm{var}(\hat{G}_E(\tau))_{can}}}{\langle
\hat{G}_E(\tau) \rangle} \approx
\frac{\frac{1}{\tau^{\frac{3}{2}}}
e^{(1-\frac{\beta}{2})\tau}}{\frac{1}{\tau} e^{(1-\beta)\tau}}
=\frac{e^{\frac{\beta \tau}{2}}}{\sqrt{\tau}},
\end{equation}
showing that the time scales are indeed much larger, as they are
now in units of $\frac{1}{\beta} \sim S$. This is not the whole
story, though. As with the toy model of the free boson, the
correct ensemble to use is the ensemble of all possible
superpositions of ground states (\ref{ground}). This will again
give an additional suppression by $e^S$ to the result computed
above, and the standard deviation to mean-ratio will behave as
\begin{equation}
\frac{\sqrt{\textrm{var}(\hat{G}_E(\tau))_{\CM_{sup}}}}{\langle
\hat{G}_E(\tau) \rangle } \approx \frac{e^{\frac{\beta
\tau}{2}}}{e^{\frac{S}{2}}\sqrt{\tau}} = \frac{e^{\frac{\beta
\tau- \frac{1}{\beta}}{2}}}{\sqrt{\tau}} ,
\end{equation}
showing that the timescale in imaginary time for distinguishing different microstates will be
$\tau \sim \frac{1}{\beta^2} \sim S^2$.   Note also that the absence of periodicity in imaginary time immediately distinguishes correlation functions computed in a microstate from correlation functions computed in the thermal ensemble.

\subsection{Lorentzian variance}
It can be readily shown that the Lorentzian variance decreases as $N$ increases, and vanishes in the large $N$ limit.   The quantity of interest for us is the standard deviation to mean ratio. In this case the ratio can be computed using (\ref{varN}) and (\ref{gravitoncorr}), and one
finds
\begin{equation}
\mathcal{F}(t)\equiv
\frac{\sqrt{\textrm{var}(\hat{G}(t))_{can}}}{\langle \hat{G}(t)
\rangle} = \frac{ \sqrt{ \sum_{n=1}^{\infty} n^2 f_n(t)^2
\frac{\cosh \beta n}{\sinh^2 \beta n}}}{\sum_{n=1}^{\infty} n
f_n(t) \frac{1}{\sinh \beta n}}, \quad \textrm{where } f_n(t) =
\left(\frac{\sin \frac{t}{2}}{n \sin \frac{t}{n}}\right)^2 \left(
1+ \frac{\sin t}{n \tan \frac{t}{n}}\right). \label{stdmeanlor}
\end{equation}
\begin{figure}
\begin{center}
\includegraphics[scale=0.6]{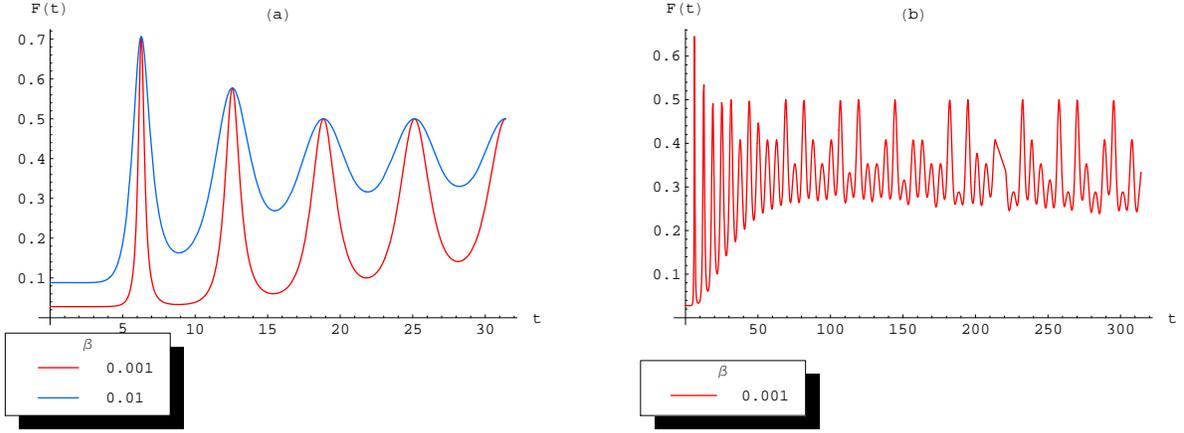}
\end{center}
\caption{\small{The standard deviation to mean ratio $\mathcal{F}(t)$ in real time.
(a): Short time scale behaviour, plotted for $\frac{1}{\beta} = 100$ (upper) and $\frac{1}{\beta} = 1000$ (lower)). (b): Long time behaviour plotted for $\frac{1}{\beta} = 1000$)}}
\label{lor}
\end{figure}
This quantity is plotted for small and large timescales in figure
\ref{lor}, which shows quasi-periodic oscillations of period
$2\pi$. To understand this behavior, we need to know which terms
contribute to the sums, and how the functions $f_n(t)$ behave.

Firstly, the terms are suppressed by $\frac{1}{\sinh \beta n}$,
and therefore for high $n$ the terms are negligible. Thus it is
enough to include only $\mathcal{O}(\frac{1}{\beta}) =
\mathcal{O}(\sqrt{N})$ first terms. Also, for small values of $n$
we can approximate $\frac{\cosh \beta n}{\sinh^2 \beta n} \approx
\frac{1}{\sinh^2 \beta n}$, showing that these terms in the two
sums in (\ref{stdmeanlor}) will scale identically as a function of
$\beta$. Thus we see that the terms that determine the behavior of
the ratio as a function of $\beta$ are the intermediate ones: $n
\lesssim \sqrt{N}$. The functions $f_n(t)$ also play an important
role and we must understand their behavior. It is easy to show
that these functions fluctuate between 0 and $\frac{1}{2}$
reaching $f_n(t) = \frac{1}{2}$ at $t = 2nm \pi$, where $m$ is any
integer, or also half integer in the case of even $n$.

\noindent{\em $\mathcal{F}$ for early times:}
Let us first analyze the behavior of the ratio around $t=0$. In
this case $f_n(t) = \frac{1}{2}$ for all $n$, and we can compute
the sums by approximating them with integrals:
\begin{equation}
\frac{\sqrt{\textrm{var}(\hat{G}(0))_{can}}}{\langle \hat{G}(0)
\rangle} \approx \frac{ \sqrt{ \frac{1}{\beta^3} \int_0^{\infty}
dx x^2 \frac{\cosh x}{\sinh^2 x}}}{\frac{1}{\beta^2}
\int_0^{\infty} dx x \frac{1}{\sinh x}} \approx 0.9 \sqrt{\beta}.
\end{equation}
This indicates that for the earliest times at high temperatures,
the fluctuations in this ensemble are small and one can't tell the
thermal state from a typical state.

\noindent{\em Peaks:}
Now let us analyze the peaks that occur at $t = 2\pi m$ for
integer $m$. At these times both numerators in $f_n(t)$ vanish, so
to get a non-zero contribution the denominators should vanish as
well. This gives the condition $\frac{2\pi m}{n} = p\pi$ for some
integer $p$, or equivalently
\begin{equation}
n = \frac{2m}{p}.
\end{equation}
For the first and highest peak $m=1$, so clearly the only
solutions are $n = 1,2$. This means that in the sums in
(\ref{stdmeanlor}) only the first two terms contribute, giving
\begin{equation}
\left. \frac{\sqrt{\textrm{var}(\hat{G}(t))_{can}}}{\langle
\hat{G} \rangle} \right|_{\textrm{first peak}} \approx
\frac{\sqrt{\frac{1}{2^2} \frac{\cosh \beta}{\sinh^2 \beta} + 2^2
\frac{1}{2^2} \frac{\cosh 2\beta}{\sinh^2 2\beta}}}{\frac{1}{2}
\frac{1}{\sinh \beta} + 2\frac{1}{2} \frac{1}{\sinh 2\beta}}
\approx \frac{\sqrt{2}}{2} \approx 0.707,
\end{equation}
which matches well with the plot in figure \ref{lor}.  Note that the height of the peak is not dependent on
the temperature $\frac{1}{\beta}$ for small $\beta$ (large $N$). The other peaks can be analyzed
using the same method.  

For later times, $t \sim \mathcal{O}(\frac{1}{\beta})$, the number
of terms that contribute at the peaks is the number of integers
that divide $2m$, which for large $m$ is generically proportional
to $\ln{m}$ \cite{dirichlet}. In this case the standard deviation
to mean ratio goes as
\begin{equation}
\mathcal{F}(t) \approx \frac{\sqrt {(\frac{1}{2})^2
\frac{1}{(\beta)^2} \times \ln \frac{1}{\beta}}}{(\frac{1}{2})
\frac{1}{\beta} \times \ln \frac{1}{\beta}} \approx
\frac{1}{\sqrt{\ln \frac{1}{ \beta}}} \approx
\frac{1}{\sqrt{\frac{1}{2} \ln N}},
\end{equation}
where $\ln \frac{1}{\beta}$ is the number of contributing terms,
all of which are roughly the same size. This shows that the height
of the peaks decreases as $N \to \infty$, although very slowly.
However, even for late times there will be many peaks that remain
finite. This is because the number of divisors of $m$ is only very
roughly $\ln m$; for instance when $m$ is prime the number of
divisors is 2, in which case the multiplicity factor is absent in
the computation above and the ratio is finite for all $\beta$.

The $O(1)$ height of the peaks in the standard-deviation to mean ratio might have suggested that the microstates can be easily distinguished from each other and from the thermal ensemble at early timescales.  If so this would contradict the finding in \cite{d1d5} that the graviton correlator is universal  and largely independent of the twist distribution for a very long period of time, and the result presented there that correlators in the basis microstates agree with the BTZ result for a time of order $1/\beta \sim \sqrt{N} \sim S$.   The potential tension is resolved, because it can be shown that in the small $\beta$ (large $N$) limit that is relevant for the validity of classical geometry, at the times $t = 2\pi m$ the mean correlator actually vanishes.  Hence although the standard deviation to mean ratio is nominally of $O(1)$ at these instants, the variation between microstates will not be measurable without very high precision.  Likewise, notice from Fig.~3 that the width in the peaks of the standard-deviation to mean ratio become narrower as $\beta$ decreases.  This indicates that in addition to precision of measurement, high temporal resolution would also be needed to resolve the differences in the correlators between different basis microstates in the ensemble $\mdiscr$.  In any case, as we will see in the ensemble of all microstates (i.e. including superpositions of the basis elements) the standard deviation to mean ratio will be enormously suppressed.

There is one more timescale that is of interest to us. As pointed
out in \cite{d1d5}, for any finite $N$ only a finite number of the
twist operators are present, and therefore there is only a finite
number of frequencies present in correlator (\ref{gravitoncorr}).
Thus the system will exhibit exact periodicity at timescales when
$t$ is of the order of the lowest common multiple of the
frequencies. This timescale was shown to increase as $e^{\sqrt{N}}
\sim e^S$, and thus the period for this system will be roughly the
Poincar\'{e} recurrence time, as expected.

\noindent{\em Suppression: }
Finally, as argued in Sec.~2, the correct ensemble to use is
$\mathcal{M}_{sup}$, the ensemble containing the superpositions of
the basis states. In this ensemble the variance is suppressed by a
factor of $e^S$ from the one computed in the ensemble
$\mathcal{M}_{bas}$, and taking this into account, we see that the
standard deviation to mean ratio computed in (\ref{stdmeanlor}) needs to divided by
$e^{\frac{S}{2}}$, making it virtually vanish. Due to the analysis
above we know that the ratio never grows
appreciably, from which we can conclude that the fluctuations in
the ensemble of superpositions are too small to be detected by any semiclassical
measurement, and therefore one cannot
hope to tell different microstates apart from each
other.

\paragraph{Summary of the D1-D5 system:} We have shown that in the D1-D5 system the individual microstates can be distinguished from the $M=0$ BTZ geometry by excursions of the correlation functions into the complex plane. Despite the entropic suppression of the variance we find that at imaginary timescales $\tau \sim S^2$ there is a distinction between the microstates and the black hole geometry. On the other hand it is virtually impossible to tell apart the states from each other in real time since the exponential suppression of the variance overwhelms the deviations in the correlation functions. The natural timescale over which we can tell apart states is expected to be $t \sim e^S$ -- the Poincar\'e recurrence time ({\it cf.}, \cite{vijaydoninteg}).

%----------------------------------------------------------------------------------
\section{Discussion}
\label{discussion}
%----------------------------------------------------------------------------------

Our principal finding is that the variances of local correlation functions computed in generic microstates of a system with entropy $S$ are suppressed by a factor of $e^{-S}$.   This is a general result arising from statistical considerations and is true both for free and interacting theories regardless of the strength of the coupling.    Our results were illustrated in two examples: the 1+1 dimensional free boson and the D1-D5 CFT which is dual to the BTZ black hole.  Applied to black holes it implies that extreme precision (and correspondingly long measurement timescales) are necessary to distinguish microstates from each other.    Thus our results suggest that even if there are non-singular, horizon free black hole microstates as proposed in \cite{fuzzball}, they are universally described by the semiclassical, coarse-grained observer in terms of the conventional black hole geometries.\footnote{In the work of \cite{liufestuccia} it was shown that perturbative supersymmetric Yang-Mills theory does not reproduce the thermal behaviour required to describe a dual black hole, and strong coupling dynamics was essential to have a field theory description of black holes.  Here we have been able to avoid that issue, because we have assumed that the Hamiltonian has the requisite gap structure and degeneracy; then the main result follows from statistical considerations.   In the description of the BTZ $M=0$ black hole using the D1-D5 CFT, the free orbifold theory already reproduced interesting aspects of the black hole physics because of the ``fractionation" of momenta in the  CFT -- this is  sufficient to produce the right gap structures.}

In our two examples we investigated whether the structure of correlation functions in imaginary time can better separate microstates from each other and from the thermal average.   Certainly the thermal average correlator differs from the answer in any individual microstate because the former must be periodic in imaginary time.  In particular, the correlation function at zero spatial separation must therefore diverge at $\tau = \beta$ in the thermal ensemble, while it is finite for a generic microstate.   Furthermore, while we have seen that the variances in the correlation functions are exponentially suppressed, making it impossible to distinguish different states in real time-scales less than the Poincar\'e time $t \sim e^S$, correlation functions grow exponentially in imaginary time, which possibly increases the differences between distinct microstates.   While it is not clear that analytically continued correlation functions are accessible to a single observer, we should emphasize that the discussion above indicates an in-principle possibility of being able to access the information and hence being able to distinguish microstates.

It is worth emphasizing that in this picture, individual microstates of fixed energy do {\it not} correspond to an eternal black hole geometry.  This is because correlators computed in a single asymptotic region of the latter are automatically periodic in imaginary time, unlike the correlators in individual microstates.   Thus, we must conclude that the canonical (fixed temperature) and microcanonical (fixed energy) ensembles in a CFT are not dual to eternal black holes as they are sometimes taken to be.      Rather, the ensemble in the CFT is dual to an ensemble of microstate spacetimes which may or may not have a description purely in geometry, but certainly do not have two asymptotic regions and imaginary time periodicity of correlators.     However, our findings imply that a {\it semiclassical} observer with access only to coarse-grained observables would be able to describe his or her findings approximately in terms of a BTZ geometry.

Finally, our analysis shows the exponential suppression of variances for finitely local correlation functions; the explicit calculations used specific local correlators that were easy to compute.  One might ask if there are other observables that can separate the microstates more easily.   For example, nonlocal observables like Wilson loops are infinite sums of local correlators and have been argued to be effective at probing the underlying states of black holes \cite{wilsonloop}.    Our finding that the variance in correlators is amplified in imaginary time has a related character because to determine the correlator everywhere in the complex plane from measurements just along the real line will require knowledge of an infinite number of derivatives.\footnote{Note, though, that it is possible to use  techniques like Pad\'e approximants to extrapolate from finite data sets to obtain some information about the analytic properties.}

\vspace{0.2 in}

\paragraph{Acknowledgments: } We have enjoyed useful discussions with Vishnu Jejjala, Per Kraus,  Hong Liu, Steve Shenker and Masaki Shigemori.   Some of the work in this paper was done at the 2006 Aspen workshop on black hole physics, the 2006 Indian String Meeting, the 2007 KITP miniprogram on the quantum nature of singularities, and under a banyan tree at the Konarak temple.  VB, BC, KL and JS were supported in part by DOE grant DE-FG02-95ER40893.    JS was supported in part by DOE grant DE-AC03-76SF00098 and NSF grant PHY-0098840.   This research was also supported in part by the National Science Foundation under Grant No. PHY99-07949.   
%We thank Atish Dabholkar for dancing lessons and an advance citation complaint.
\appendix

\section{Ensembles with fixed energy expectation values}
\Label{realisticth}

We have considered ensembles of states in which the energy
eigenvalues fall between $E$ and $E + \Delta E$.  One might have
considered an ensemble of states in which only the expectation
values are so bounded, i.e.
\begin{equation}
\CM_{exp} = \left\{ \ |\psi\rangle \ : \  E \leq \, \langle \,
\psi | H | \psi \, \rangle \, \leq E + \Delta E \ \right\} \, .
\Label{expensemble}
\end{equation}
The states $|\psi\rangle$ could then be superpositions of energy
eigenstates with arbitrarily high eigenvalues.    For example, we
could take
\begin{equation}
|\psi\rangle = (1- \epsilon)^{1/2} \, |0\rangle + \epsilon^{1/2}
\,|s \rangle \ \ \Longrightarrow \ \ \langle\psi | \,H \,|\psi
\rangle = \epsilon \, e_s = E
\end{equation}
with $e_s \gg E$ and $\epsilon = E/e_s$. In general if we take
$|\psi\rangle = \sum_s c_s |s\rangle$ to be an element of
$\CM_{exp}$ written as a sum of energy eigenstates, then we
require that
\begin{equation}
E \leq \sum_s |c_s|^2 \, e_s \leq E+ \Delta E \, .
\end{equation}
Since this is not a linear constraint on the coefficients $c_s$,
it is not possible to write a linear basis of states for all
elements of the ensemble $\CM_{exp}$;  i.e. they do not form a
Hilbert space.     Of course the superposition coefficients of the
eigenstates with energies much bigger than $E$ will have to be
small, but observables such as $H^k$ for large $k$ will be
increasingly sensitive to the energy components of states lying
outside the range $E \leq e_s \leq E + \Delta E$.

In view of this, the essential finding -- that the variance over
$\mcont$ is suppressed -- will apply to this ensemble when probe
operators satisfy a Lipschitz condition with respect to energy
expectations:
\eqn{lipschitz}{|\langle\,\psi\,|\,\mathcal{O}\state{\psi} -
\langle\,\phi\,|\,\mathcal{O}\state{\phi}| < L\,
|\langle\,\psi\,|\,H\state{\psi} -
\langle\,\phi\,|\,H\state{\phi}|} for some constant $L$. To see
this, consider a family of states $|E\rangle$, parametrized by the
real scalar $b$:
\eqn{stateE}{\state{E} = \sqrt{\frac{b-E}{b-a}} \state{a} +
\sqrt{\frac{E-a}{b-a}}\state{b} ,} with:
\eqn{stateab}{\eqalign{\langle\, E\, | \,H \state{E} & = E \cr H
\state{a} & = a \state{a} \cr H \state{b}  & = b \state{b}\, .}}
In the above, the fixed constant $a$ is an energy close to but
lower than $E$
\eqn{conditionsa} {E -\Delta E < a < E} while the parameter $b$
must be greater than $E$ but is otherwise unconstrained. Then:
\eqn{oexpectation}{\eqalign{\langle\, E\, | \, \mathcal{O}
\state{E} & = \frac{b-E}{b-a} \langle\, a\, | \, \mathcal{O}
\state{a} + \frac{E-a}{b-a} \langle\, b\, | \, \mathcal{O}
\state{b} \cr & = \langle\, a\, | \, \mathcal{O} \state{a} +
\frac{E-a}{b-a} \left(\langle\, b\, | \, \mathcal{O} \state{b} -
\langle\, a\, | \, \mathcal{O} \state{a}\right) \cr & < \langle\,
a\, | \, \mathcal{O} \state{a} + L\, \Delta E\, .}}
In this way, the Lipschitz condition will bound discrepancies in
means and variances between the ensembles $\CM_{exp}$ and
$\CM_{sup}$ to $\mathcal{O}(\Delta E)$.

%----------------------------------------------------------------------------------
\section{Derivation of the ``thermal state''}
\Label{typmicro}
%----------------------------------------------------------------------------------

The occupation numbers $N_n$, which characterize elements of the
ensemble $\mdiscr$, satisfy $\sum n N_n = E$ and therefore admit a
representation in terms of Young tableaux $\mu$. Such tableaux are
conveniently described in a coordinate system, where the
horizontal axis $y$ spans the indices of the bosonic oscillators,
while the abscissa $x$ measures the cumulative population of all
the oscillators from infinity down to $y(x)$:
\eqn{identification0}{ x(y) = \int_y^\infty N_n \, dn \, .}
Using canonical ensemble methods, Vershik showed \cite{vershik}
that for large $E$, almost all states (elements of $\mdiscr$) lie
close to the limit shape given by:
\eqn{vershiklimit}{ \exp{(-x\sqrt{\zeta(2)/E})} +
\exp{(-y\sqrt{\zeta(2)/E})} = 1 \, .}
The curve \vershiklimit\ may be thought of as the average of the
ensemble $\mdiscr$, and as such it should be identified with the
thermal state \typicalN . This is accurate for regimes where the
canonical and the microcanonical treatments agree, i.e. for $x, \,
y \approx \sqrt{E}$, which is where most of the entropy lies. The
occupation numbers obtained from the limit curve take the form
\eqn{identification1}{ N_n = - \frac{dx}{dy} \biggr|_{y=n} \; = \;
{1\over\exp{(n\sqrt{\zeta(2)/E})}-1} \, ,}
leading to the identification:
\eqn{identification2}{ \beta = \sqrt{\frac{\zeta(2)}{E}} \, .}
The entropy, given by the asymptotic total number of partitions of
$E$ due to Hardy and Ramanujan \cite{hardy}, takes the form:
\eqn{entropy} {S = \frac{2\, \zeta(2)}{\beta} +
\mathcal{O}(\log{\beta}) \, .}

%----------------------------------------------------------------------------------
\section{Taxonomy of possible configurations}
\Label{taxonomy}
%----------------------------------------------------------------------------------

We collect in this appendix the various states and ensembles that are used in the main discussion. Since one is generically interested in telling apart microstates from each other and from the black hole it is useful to keep the subtle distinctions described below in mind while referring to various macroscopic configurations.

\paragraph{Ensembles:}
\begin{enumerate}
\item Canonical or thermal ensemble: This is the ensemble of states in the Hilbert space with the probability of finding an energy eigenstate of energy $E$ is given by $e^{-\b E}$.
This ensemble is best thought of for measurement purposes in the quantum cosmological sense. Each measurement gives us the eigenvalues of the probe, subject to the fact that the intrinsic probability distribution of states is given by the canonical distribution.
\item Microcanonical basis ensemble, $\mdiscr$: This is the ensemble defined in (\ref{Eeigenstates}), where we restrict attention to the basis of energy eigenstates (which for the cases of interest are also number operator eigenstates), with each element being equally probable in the ensemble. The ensemble has $e^S$ elements.
\item Microcanonical superposition ensemble, $\mcont$: This is the ensemble defined in (\ref{allstates}), where we allow arbitrary superpositions of energy eigenstates, appropriately normalized.
\end{enumerate}

\paragraph{States:}
\begin{enumerate}
\item Hartle-Hawking state: This is a pure entangled state that is constructed in the tensor product of two copies of the Hilbert space. The entanglement is fine tuned so that we reproduce the average values of the canonical ensemble after tracing over one of the Hilbert spaces. This is the description of the eternal Schwarzschild-AdS black hole.
\item ``Thermal state'': The ``thermal state'' a formal construct; it doesn't belong to any of the  ensembles described above which reproduces thermal averages. This is the state whose derivation was explained in Appendix ~\ref{typmicro}.\footnote{It can be thought of as a superposition of energy eigenstates with the coefficients given by the thermal distribution. It is also likely that the ``thermal state'' is the result of the tracing procedure in the Hartle-Hawking state. Hence, it can also be thought of as the description of the black hole as viewed from a single Hilbert space. }
\item Typical microstate: The typical microstate is an element of $\mcont$ whose occupation numbers are tuned by choice of the superposition coefficients to come arbitrarily close to mimic the ``thermal state'', although only up to the energy of the ensemble $\mcont$.
\item Typical basis microstates: These states are elements of $\mdiscr$, with integral occupation numbers which lie close to the typical microstate. In the geometric picture they are usually associated with the microstate geometries of \cite{fuzzball}.
\end{enumerate}

%=============================================================

%=============================================================

%\bibliography{thtyp_refs}
%\bibliographystyle{utphys}

%_____________________________________________________________

\end{document}